\documentclass[a4paper,twocolumn,superscriptaddress,11pt,accepted=2018-12-18]{quantumarticle}
\pdfoutput=1
\usepackage[utf8]{inputenc}
\usepackage[english]{babel}
\usepackage[T1]{fontenc}
\usepackage{amsmath}
\usepackage{hyperref}
\usepackage[usenames, dvipsnames]{color}
\usepackage{graphicx}
\usepackage{gensymb}
\usepackage{subfigure}
\usepackage{pifont}
\usepackage[braket, qm]{qcircuit}
\usepackage{listings}
\usepackage{mathtools}
\usepackage{float}

\begin{document}

\title{Exact Ising model simulation on a quantum computer}
\date{\today}
\author{Alba Cervera-Lierta}
\affiliation{Barcelona Supercomputing Center (BSC), Barcelona, Spain}
\affiliation{Institut de Ci\`encies del Cosmos, Universitat de Barcelona, Barcelona, Spain}
\email{a.cervera.lierta@gmail.com}
\orcid{0000-0002-8835-2910}

\maketitle

\begin{abstract}
We present an exact simulation of a one-dimensional transverse Ising spin chain with a quantum computer. We construct an efficient quantum circuit that diagonalizes the Ising Hamiltonian and allows to obtain all eigenstates of the model by just preparing the computational basis states. With an explicit example of that circuit for $n=4$ spins, we compute the expected value of the ground state transverse magnetization, the time evolution simulation and provide a method to also simulate thermal evolution. All circuits are run in IBM and Rigetti quantum devices to test and compare them qualitatively. 
\end{abstract}

\section{Introduction}

In recent years the quantum computing has dived fully into the experimental realm. Control of quantum systems has improved so much that quantum computing devices have become a near term reality. These experimental advances rely on some criteria proposed in the 2000's by DiVicenzo \cite{DiVicenzo}: scalable physical system to characterize the qubits, simple fiducial qubit state initialization, long coherence times (longer than the gate implementation times), universal set of quantum gates and qubit-specific measurement capability. Although there are some candidates that can fulfill the first criterion, the field is still in an early stage of development of this technology, where the improvement of qubits control is crucial to accomplish the others.

Private companies have also joined the field. Since 2016, IBM offers cloud based quantum computation platform \cite{IBM}. Any user can run quantum algorithms on their two five qubits devices, their 16 qubits device, and their 20 qubits device which is available for hubs and partners. It is not the only company that has launched this kind of service: Rigetti Computing also allows the use of its 19-qubits device on the cloud \cite{Rigetti}. Although both companies are betting for superconducting qubits, their respective device characterization is not the same: basic gate sets and qubits connectivity are some of the differences. As more quantum devices are appearing, it is important to find some methods to test their quality when running sophisticated quantum algorithms.
%to figure out their quality and have a way to compare them when a 

Several approaches on computer's \emph{quantumness} have already been tried. The first published article using an IBM device tested the violation of Bell inequalities by more than two qubits (Mermin inequalities) \cite{Dani}, or a recent article tests if the 16-qubit IBM device can be fully entangled by generating graph states \cite{Ent}. Other works tried to exploit different few qubit experiments, such as error correcting codes and quantum arithmetics \cite{Devitt}.

On the other hand, the community has not forgot Feynman's original aim for the proposal of construction of a quantum computer \cite{Feynman}: the simulation of quantum systems. Many classical techniques have been developed in that direction, for instance quantum Monte Carlo methods \cite{QMM1,QMM2,QMM3} or tensor networks algorithms \cite{TN}. However, the first suffer from the well-known sign problem  and the second are only efficient for slightly entangled systems \cite{Vidal}. In the end, very strongly correlated quantum systems, such as those displaying frustration, will need a quantum computer to be efficiently simulated \cite{original}. There are some works that propose quantum algorithms to construct arbitrary Slater determinants, both in one and two dimensions, to simulate the dynamics of the ground state of fermionic hamiltonians, in particular the Hubbard model \cite{slater1,slater2}. Other proposals introduce the concept of \emph{compressed quantum computation}, i.e. simulation of $n$-spin chain using $\log n$ qubits \cite{Kraus}. This method has been tested in one of the IBM's quantum computers also simulating the transverse magnetization of the Ising model \cite{Dani2}: the main difference respect to the work proposed in that paper is we have access to the whole energy spectrum, which allows us to simulate time and temperature evolution as well.

In this work, we implement a four-qubit experiment that could be interesting both as a proposal for testing and comparing devices quality and for its implications in condensed matter physics. We perform the \emph{exact} simulation of a spin chain proposed in Ref.\cite{Latorre} with an Ising-type interaction. The Ising model is one of the most famous exactly solvable models, i.e. those models that are integrable. Actually, the steps to find a quantum circuit that diagonalizes the Ising Hamiltonian follow the same strategy than the analytical solution of the model. Therefore, the method can be extended to other integrable models like the Kitaev-honeycomb model, which a circuit has already been proposed \cite{Orus}. As we are performing an exact simulation, we have access to the whole spectrum and not only to ground state: time evolution and thermal states can be simulated exactly as well. 
This provides a new approach in quantum simulation if an exact circuit is found for those non trivial models, such as Heisenberg model, which have an ansatz to be solved. In particular, for one-dimensional spin chains, the Bethe ansatz \cite{Bethe} is the most successful method and several proposals exist to simulate and extend it to two-dimensions using tensor network techniques \cite{Verstraete}. As the one-dimensional Ising model has analytic solutions for arbitrary number of spins and the circuit proposed in this paper can be efficiently generalized to larger number of qubits, the methods outlined in this work can be used to benchmark a quantum computer by seeing how this compares against known solutions.

The paper is structured as follows. In section \ref{sec:circuit} we describe the method proposed in Ref.\cite{Latorre} to construct an efficient circuit that diagonalizes the Ising Hamiltonian: the number of gates scales as $n^2$ and the circuit depth as $n\log n$. In section \ref{sec:time} we explain briefly the basic concepts of time evolution in quantum mechanics and give a specific example to be simulated using the circuit derived in the previous section. In section \ref{sec:thermal}, we propose two methods to simulate the expected value of an operator for finite temperature. Section \ref{sec:computer} summarizes the properties of the three devices used for this work, two from IBM and one from Rigetti, and in section \ref{sec:results} we present the results of ground state transverse magnetization and the time evolution of $|\uparrow\uparrow\uparrow\uparrow\rangle$ state and compare the three devices according to them. Finally, the conclusions are exposed in section \ref{sec:conclusion}.

\section{Quantum circuit for the Ising Hamiltonian}\label{sec:circuit}

Let's consider the existence of a quantum circuit that \textit{disentangles} a given Hamiltonian and transforms its entangled eigenstates into product states. This circuit will be represented by an unitary transformation $U_{dis}$
\begin{equation}
\widetilde{\mathcal{H}}=U_{dis}^{\dagger}\mathcal{H}U_{dis},
\end{equation}
where $\mathcal{H}$ is the model Hamiltonian and $\widetilde{\mathcal{H}}$ is a noninteracting Hamiltonian that can be written as $\widetilde{\mathcal{H}}=\sum_{i}\epsilon_{i}\sigma_{i}^{z}$. This diagonal Hamiltonian contains the energy spectrum $\epsilon_{i}$ of the original one and its eigenstates correspond to the computational basis states. Then, we will have access to the whole spectrum of the model by just preparing a product state and applying $U_{dis}$.

There exist infinite $U_{dis}$ gates for a given Hamiltonian. In general, to find these disentangling unitaries will be a hard task, probably as hard as finding a method to diagonalize analytically the Hamiltonian. However, for some models we can follow a kind of recipe to construct a disentangling gate. For the case of Ising Hamiltonian, the steps to obtain the $U_{dis}$ quantum gate are based on the analytical solution of the model \cite{Lieb,Katsura}: {\sl i)} Implement the Jordan-Wigner transformation to map the spins into fermionic modes. {\sl ii)} Perform the Fourier transform to get fermions to momentum space. {\sl iii)} Perform a Bogoliubov transformation to decouple the modes with opposite momentum. Thus, the construction of the disentangling gate can be done by pieces:
\begin{equation}
U_{dis}=U_{JW}U_{FT}U_{Bog}.
\end{equation}
In the following subsections, we derive the quantum gates needed to implement the above transformation.

\subsection{Jordan-Wigner transformation}

Let's start with the antiferromagnetic Ising Hamiltonian with transverse field
\begin{equation}
\mathcal{H}=\sum_{i=1}^{n}\sigma_{i}^{x}\sigma_{i+1}^{x} + \sigma_{1}^{y}\sigma_{2}^{z}\cdots\sigma_{n-1}^{z}\sigma_{n}^{y} +\lambda\sum_{i=1}^{n}\sigma_{i}^{z},
\end{equation}
where $\lambda$ is the transverse field strength. The second term has been added to cancel the periodic boundary term, $\sigma_{n}^{x}\sigma_{1}^{x}$, after the Jordan-Wigner transformation in order to solve the system as it was infinite. This modified Hamiltonian will have finite size effects that become negligible as $n$ grows. 

The Jordan-Wigner transformation corresponds to transform the spin operators $\mathbf{\sigma}$ into fermionic modes $c$ \cite{JW}:
\begin{equation}
c_{j}=\Bigg(\prod_{l<j}\sigma_{l}^{z}\Bigg)\frac{\sigma_{j}^{x}+i\sigma_{j}^{y}}{2}, \  c_{j}^\dagger=\frac{\sigma_{j}^{x}-i\sigma_{j}^{y}}{2}\Bigg(\prod_{l<j}\sigma_{l}^{z}\Bigg),
\end{equation}
where $c_{j}$ and $c_{j}^{\dagger}$ are the fermionic annihilation and creation operators acting on the vacuum $|\Omega_{c}\rangle$, $c_{i}|\Omega_{c}\rangle=0$, and following the anticommutation rules $\{c_{i},c_{j}\}=0$ and $\{c_{i},c_{j}^\dagger\}=\delta_{ij}$. After this transformation the Hamiltonian reads
{\medmuskip=-1mu
\thinmuskip=0mu
\thickmuskip=0mu
\begin{equation}
\mathcal{H}_{c}=\frac{1}{2}\sum_{i=1}^{n}\big(c_{i}^{\dagger}c_{i+1}+c_{i+1}^{\dagger}c_{i}+c_{i}c_{i+1}+ c_{i}^{\dagger}c_{i+1}^{\dagger}\big)+\lambda\sum_{i=1}^{n}c_{i}^{\dagger}c_{i}.
\end{equation}}

In terms of the wave function,
\begin{eqnarray}
|\psi\rangle&=&\sum_{i_{1},\cdots,i_{n}=0,1}\psi_{i_{1}\cdots i_{n}}|i_{1}\cdots i_{n}\rangle \nonumber\\
&=&\sum_{i_{1},\cdots,i_{n}=0,1}\psi_{i_{1}\cdots i_{n}} (c_{1}^{\dagger})^{i_{1}}\cdots (c_{n}^{\dagger})^{i_{n}}|\Omega_{c}\rangle.
\end{eqnarray}
Notice that the coefficients $\psi_{i_{1}\cdots i_{n}}$ do not change. Then it will not be necessary to implement any gates on the quantum register to perform this transformation. However, for now on we should take into account we are dealing with fermionic modes, so any swap between two occupied modes will carry a minus sign. In terms of quantum gates, this is translated into the use of {\sl fermionic} SWAP gate (fSWAP) each time we exchange two modes:
\begin{equation}
\mathrm{fSWAP}=\left(\begin{array}{cccc}
1&0&0&0\\0&0&1&0\\0&1&0&0\\0&0&0&-1
\end{array}\right),
\label{fSWAP}
\end{equation}
which corresponds with the usual SWAP gate followed or preceded by a controlled-Z gate (see appendix A).

\subsection{Fourier Transform}

The next step to solve the Ising model consists on getting the fermionic modes to momentum space using the well-known quantum Fourier transform
\begin{equation}
%b_{k}=\frac{1}{\sqrt{n}}\sum_{j=1}^{n}\exp\left(i\frac{2\pi j}{n}k\right)c_{j}, \quad
b_{k}^{\dagger}=\frac{1}{\sqrt{n}}\sum_{j=1}^{n}\exp\left(i\frac{2\pi j}{n}k\right)c_{j}^{\dagger}, \ k=-\frac{n}{2}+1,\cdots,\frac{n}{2}.
\end{equation}

For $n=2^m$ for some integer $m$, this transformation can be implemented with a log-depth circuit and using at most two-body quantum gates. This method is called {\sl fast Fourier transform} and consists in two parallel Fourier transformations over $n/2$ sites, the even and the odd sites \cite{FFT}:
\begin{eqnarray}
%\mathclap{
\sum_{j=0}^{n-1}e^{\frac{2\pi i k}{n}j}c_{j}^{\dagger}
&=&\sum_{j'=0}^{\frac{n}{2}-1}e^{\frac{2\pi i k}{n/2}j'}c_{2j'}^{\dagger}+e^{\frac{2\pi i k}{n}}
%\sum_{j'=0}^{\frac{n}{2}-1}
e^{\frac{2\pi i k}{n/2}j'}c_{2j'+1}^{\dagger}. \nonumber\\%{\color{white}xxx}}
\end{eqnarray}
To implement such a transformation we need a combination of a two-qubit gate, a `beam-splitter' $F_2$, and one-qubit gate, the `phase-delay' $\omega_{n}^{k}$, which applies the so-called twiddle-factor $e^{2\pi i k/n}$:
\arraycolsep=1.8pt\def\arraystretch{1.2}
\begin{equation}
F_{2}=\left(\begin{array}{cccc} 1&0&0&0\\ 0& \frac{1}{\sqrt{2}}& \frac{1}{\sqrt{2}} &0\\
0& \frac{1}{\sqrt{2}}& -\frac{1}{\sqrt{2}} &0\\ 0&0&0& -1\end{array}\right), \  \omega_{n}^{k}=\left(\begin{array}{cc}1&0\\0& e^{\frac{2\pi ik}{n}} \end{array}\right),
\end{equation}
where the fermionic anticommutation relation has been taken into account in the $-1$ element of the $F_{2}$ matrix. 

All together, the Fourier transform gate becomes
\arraycolsep=1.8pt\def\arraystretch{1.2}
\begin{equation}
F^{n}_{k}=\left(\begin{array}{cccc} 1&0&0&0\\ 0& \frac{1}{\sqrt{2}}& \frac{e^{\frac{2\pi ik}{n}}}{\sqrt{2}} &0\\
0& \frac{1}{\sqrt{2}}& -\frac{e^{\frac{2\pi ik}{n}}}{\sqrt{2}} &0\\ 0&0&0& -e^{\frac{2\pi ik}{n}}\end{array}\right).
\label{F_gate}
\end{equation}
The appendix shows the explicit decomposition of this gate.

After the Fourier transformation, the Hamiltonian becomes
\begin{multline}
\mathcal{H}_{b}=\sum_{k=-n/2+1}^{n/2}\left[2\left(\lambda-\cos\left(\frac{2\pi k}{n}\right)\right)b_{k}^{\dagger}b_{k} \right.\\
\left. +i\sin\left(\frac{2\pi k}{n}\right)\left(b_{-k}^{\dagger}b_{k}^{\dagger}+b_{-k}b_{k}\right)\right],
\end{multline}
which it is not diagonal yet as modes with opposite momentum are still coupled.

\subsection{Bogoliubov transformation}

The last step will consist on finding a transformation which mixes the two modes according to
\begin{eqnarray}
a_{k}&=&u_{k}b_{k}+iv_{k}c_{-k}^{\dagger},  \nonumber\\
a_{k}^{\dagger}&=&u_{k}c_{k}^{\dagger}-iv_{k}c_{-k}.
\end{eqnarray}
This will be implemented by a two-qubit gate which acts over qubits that represent opposite momenta. For the case of Ising model, this gate is
\begin{eqnarray}
&B_{k}^{n}=\left(\begin{array}{cccc}\cos\left(\frac{\theta_{k}}{2}\right)&0&0& i\sin\left(\frac{\theta_{k}}{2}\right)\\0&1&0&0\\0&0&1&0\\i\sin\left(\frac{\theta_{k}}{2}\right)&0&0&\cos\left(\frac{\theta_{k}}{2}\right)\end{array}\right), &\nonumber\\ &\theta_{k}=\arccos\left(\frac{\lambda-\cos\left(\frac{2\pi k}{n}\right)}{\sqrt{\left(\lambda-\cos\left(\frac{2\pi k}{n}\right)\right)^2+\sin^2\left(\frac{2\pi k}{n}\right)}}\right),&
\label{B_gate}
\end{eqnarray}
and its decomposition in basic gates is shown in the appendix.

Then, we have finally arrived to the diagonal Hamiltonian:
\begin{equation}
\widetilde{\mathcal{H}}=\mathcal{H}_{a}=\sum_{k=-n/2+1}^{n/2}\omega_{k}a_{k}^{\dagger}a_{k},
\end{equation}
where $\omega_{k}=\sqrt{\left(\lambda-\cos\left(\frac{2\pi k}{n}\right)\right)^2+ \sin^2\left(\frac{2\pi k}{n}\right)}$.

\subsection{$n=4$ spin chain}

\begin{figure}[t!]
\centering
\[
\Qcircuit @C=0.5cm @R=.5cm {
& \multigate{1}{B_{1}^{\dagger}} & \multigate{1}{F_{1}^{\dagger}}  & \qw & \qw & \multigate{1}{F_{0}^{\dagger}} & \qw & \qw & \qw \\
& \ghost{B_{1}^{\dagger}} & \ghost{F_{1}^{\dagger}} & \qw & \link{1}{-1} & \ghost{F_{0}^{\dagger}} & \qw & \link{1}{-1} &  \qw \\
& \qw & \multigate{1}{F_{0}^{\dagger}}  & \qw & \link{-1}{-1} & \multigate{1}{F_{0}^{\dagger}} & \qw & \link{-1}{-1} &  \qw \\
& \qw & \ghost{F_{0}^{\dagger}}  & \qw & \qw & \ghost{F_{0}^{\dagger}} & \qw & \qw & \qw 
  \gategroup{2}{4}{3}{5}{.0em}{-}\gategroup{2}{7}{3}{8}{.0em}{-}
  }
\]
\caption{
Quantum circuit that transforms computational basis states into transverse Ising eigenstates. The two qubit gates $F_{1}^\dagger$ and $F_{0}^\dagger$ apply the inverse Fourier transform and the $B_{1}^\dagger$ with the inverse Bogoliubov transformation. Gates represented with crosses correspond with the fSWAP gates that take care of the fermion anticommutation relations and can be removed depending on the connectivity of the quantum chip.
\vspace{-0.3cm}}
\label{Fig:circuit}
\end{figure}
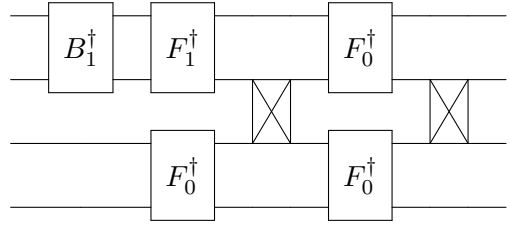

The explicit circuit for a $n=4$ chain is shown in Figure \ref{Fig:circuit}. First, we prepare the initial state as the ground state for the diagonal Hamiltonian $\widetilde{\mathcal{H}}$:
\begin{equation}
|gs\rangle=\left\{\begin{array}{c}|0000\rangle \ \mathrm{for} \ \lambda>1, \\
|0001\rangle \ \mathrm{for} \ \lambda<1. \end{array}\right. 
\end{equation}

The circuit strategy consists in undoing the steps that diagonalize the Ising Hamiltonian. Thus we first undo Bogoliubov transformation by applying $(B_{k}^{n})^\dagger$ gates, followed by undoing the Fourier transform using the $(F_{k}^{n})^\dagger$ gates and finally undo the Jordan-Wigner transformation which, fortunately, does not need from any gate as has been explained in the previous section.

For $n=4$, the Bogoliubov modes are $\pm 3\pi/2$ and $\pm \pi/2$, so we need two Bogoliubov gates. Notice that we have removed the $B_{0}^\dagger$ gate from the circuit of Figure \ref{Fig:circuit}; this gate corresponds with the identity for $\lambda>1$ and exchange qubits in the same state for $\lambda<1$, i.e. $|00\rangle\rightarrow -i|11\rangle$ and $|11\rangle\rightarrow -i|00\rangle$. As the initial state for $\lambda<1$ is the $|0001\rangle$ and $B_{0}^\dagger$ is applied over the last two qubits, it does not affect this state and we can avoid it. If we want to obtain an excited state which eigenstate in the diagonal basis contains $|00\rangle$ or $|11\rangle$ states, then we should only apply bit flip gates over the last two qubits to implement the $B_{0}^\dagger$ gate.

The circuit shown in  Figure \ref{Fig:circuit} also contains fSWAP gates represented with crosses. These will be necessary if even and odd qubits are not physically connected and, as much, they will increase the total number of gates in $n^2$. We can eliminate them if the implementation is done in the ibmqx5 device, which allow us to save up to 16 gates of depth according to IBM gate set, but they are indispensable for the implementation in the other IBM device, ibmqx4, as well as in Rigetti's 19-qubit chip. 

\newpage
\section{Time evolution}\label{sec:time}

Once we have the $U_{dis}$ circuit, we have access to the whole Ising spectrum by only implementing this gate over the computational basis states. This allows us to perform \emph{exactly} time evolution, where the characterization of all states is needed. 

The time evolution of a given state driven by a time-independent Hamiltonian is described using the time evolution operator $U(t)\equiv e^{-it\mathcal{H}}$:
\begin{equation}
|\psi(t)\rangle=U(t)|\psi_{0}\rangle=\sum_{i}e^{-it \epsilon_{i}}|E_{i}\rangle\langle E_{i}|\psi_{0}\rangle,
\label{eq:time_state}
\end{equation}
where $|\psi_{0}\rangle$ is the initial state and $\epsilon_{i}$ are the energies of the Hamiltonian states $|E_{i}\rangle$. Then, if $|\psi_{0}\rangle$ is an eigenstate of $\mathcal{H}$ there is no change in time (steady state) and therefore the expected value of an observable $\mathcal{O}$ will be constant in time. On the contrary, and if $[\mathcal{H},\mathcal{O}]\neq 0$, the expected value will show an oscillation in time given by
%{\medmuskip=0mu
%\thinmuskip=0mu
%\thickmuskip=0mu
\begin{equation}
\langle\mathcal{O}(t)\rangle=\sum_{i,j}e^{-it(\epsilon_{i}-\epsilon_{j})} \langle\psi_{0}|E_{j}\rangle\langle E_{j}|\mathcal{O}|E_{i}\rangle\langle E_{i}|\psi_{0}\rangle.
\label{eq:time_obs}
\end{equation}%}

We can take advantage from the fact that the eigenstates of the non-interacting Hamiltonian $\widetilde{\mathcal{H}}$ are the computational basis states and, as we have solved the model, we also know all energies $\epsilon_{i}$. Then, it is straightforward to construct the time evolution of a given state $|\psi_{0}\rangle$ by only expressing it in the computational basis and adding the corresponding factors $e^{-it\epsilon_{i}}$. After that, we only need to implement $U_{dis}$ gate over this state to obtain the time evolution driven by the Ising Hamiltonian.

As example, we compute the time evolution of the expected value of transverse magnetization. We take all spins aligned in the positive $z$ direction as initial state, i.e. $|\uparrow\uparrow\uparrow\uparrow\rangle$, which in the computational basis is the $|0000\rangle$ state. First, we have to express this state in the $\widetilde{\mathcal{H}}$ basis, which using $U_{dis}^{\dagger}$ become
\begin{equation}
|\psi_{0}\rangle=U_{dis}^{\dagger}|0000\rangle=\cos\phi|0000\rangle+i\sin\phi|1100\rangle,
\end{equation}
with $\phi=\arccos(\lambda/\sqrt{1+\lambda^2})/2$. Then, we apply the time evolution operator  to obtain $|\psi(t)\rangle$:
\begin{equation}
|\psi(t)\rangle=
%e^{-it(-2(\lambda+\sqrt{1+\lambda^2}))}\cos\phi|0000\rangle+ie^{-it2(\lambda-\sqrt{1+\lambda^2})}\sin\phi|1100\rangle =
\left(\cos\phi|00\rangle+ ie^{4it\sqrt{1+\lambda^2}}\sin\phi|11\rangle\right)\otimes|00\rangle.
\label{eq:time_Ising}
\end{equation}
%This state preparation is described in Appendix \ref{app:gates}.

Analytically, 
\begin{equation}
\langle\sigma_{z}\rangle=\frac{1+2\lambda^2+\cos\left(4t\sqrt{1+\lambda^2}\right)}{2+2\lambda^2},
\end{equation}
from which we can obtain the expected value of transverse magnetization, $M_{z}=\frac{1}{2}\langle\sigma_{z}\rangle$. 

\section{Thermal simulation}\label{sec:thermal}

\begin{figure}[t!]
\centering
\includegraphics[width=0.9\linewidth]{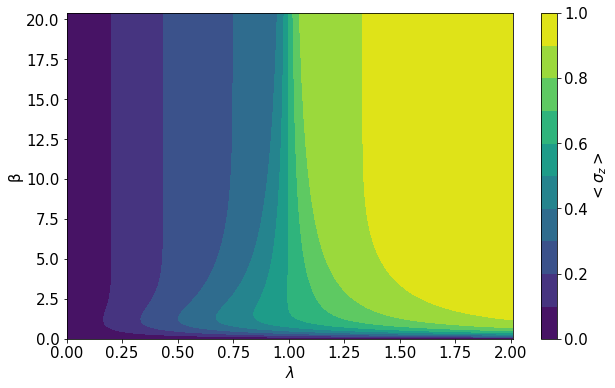}
\caption{Transverse magnetization of $n=4$ Ising spin chain as a function of temperature $\beta=1/(k_B T)$ and transverse field $\lambda$. The system undergoes a transition, from antiferromagnetic to paramagnetic, as $\lambda$ increases. For zero temperature ($\beta\rightarrow\infty$) the transition point is located at $\lambda=1$, whereas for finite temperature there is a quantum critical region around $\lambda=1$.}
\label{Fig:thermal}
\end{figure}

When a quantum system is exposed to a heat bath its density matrix at thermal equilibrium is characterized by thermally distributed populations of its quantum states following a Boltzmann distribution:
\begin{eqnarray}
\rho(\beta)= \frac{e^{-\beta\mathcal{H}}}{\mathcal{Z}}=\frac{1}{\mathcal{Z}}\sum_{i}e^{-\beta\epsilon_{i}}| E_{i}\rangle\langle E_{i}|,
\end{eqnarray}
where $\beta=1/(k_{B} T)$, $\mathcal{Z}=\sum_{i}e^{-\beta\epsilon_{i}}$ is the partition function and $\epsilon_{i}$ and $|E_{i}\rangle$ are the energies and eigenstates of the Hamiltonian $\mathcal{H}$. The expected value of some operator $\mathcal{O}$ for finite temperature is computed as
\begin{eqnarray}
\langle\mathcal{O}(\beta)\rangle=\mathrm{Tr}[\mathcal{O}\rho(\beta)]=\frac{1}{\mathcal{Z}}\sum_{i}e^{-\beta\epsilon_{i}}\langle E_{i}|\mathcal{O}|E_{i}\rangle.\nonumber\\
\end{eqnarray}

Simulate thermal evolution according to Ising Hamiltonian is, again, straightforward once we have $U_{dis}$ gate, since it consists on preparing the corresponding state in the $\widetilde{\mathcal{H}}$ basis and apply $U_{dis}$ circuit. In the case of thermal evolution, $|E_{i}\rangle$ states are the states of the computational basis, so no further gates are needed to initialize qubits apart from the corresponding combination of $X$ gates.

At that point, we can perform an exact simulation or sampling. In the first case, we run the circuit to obtain the expected value of the observable taking as initial state all states in the computational basis and average them with their corresponding energies. This is done classically once we have the expected values of each state.

On the other hand, we can perform a more realistic simulation by sampling all states according to Boltzmann distribution. First, we need to prepare classically a random generator that returns one of the computational states following the distribution $e^{-\beta\epsilon_{i}}$. Then, we run the circuit many times and compute the expected value of the operator by preparing as initial state the one returned by the generator each time.

The first method demands more runs of the experiment, $N\times 2^n$, needed for the computation of each expected value. No statistical errors come from the averaging part, as it is done classically. For the second method, with only $N$ runs we will obtain a value for the observable with a statistical error of $1/\sqrt{N}$.

For $n=4$ the transverse magnetization can be computed analytically and it is shown in Figure \ref{Fig:thermal}.
%{\medmuskip=-2mu
%\thinmuskip=0mu
%\thickmuskip=0mu
%\begin{equation}
%\langle\sigma_{z}\rangle=\frac{1}{2}\Big(\frac{\sinh(2\beta\lambda)}{\cosh(2\beta)+\cosh(2\beta\lambda)}+ \frac{\lambda\tanh(\beta+\sqrt{1+\lambda^2})}{\sqrt{1+\lambda^2}}\Big).
%\end{equation}}
%The phase diagram is very clear with magnetization, as it is shown in Fig.\ref{Fig:thermal}. 
At zero temperature, i.e. $\beta\rightarrow\infty$, the system undergoes to phase transition in the thermodynamic limit, i.e. $n\rightarrow\infty$, from antiferromagnetic to paramagnetic, at the corresponding critical point of $\lambda=1$. As temperature increases ($\beta$ decreases), the system have a critical region around $\lambda=1$ until the temperature is high enough to disorder all spins, independently of the transverse field strength (limit $\beta\rightarrow 0$) \cite{Th}.

\section{Implementation on a quantum computer}\label{sec:computer}

\subsection{IBM Quantum Experience}

\begin{figure}
\centering  
%\subfigure%[$ \ $ibmqx2 ]%`Sparrow']
%{\includegraphics[width=0.25\columnwidth]{ibmqx2.png}}
\subfigure%[$ \ $ibmqx4]
{\includegraphics[width=0.22\linewidth]{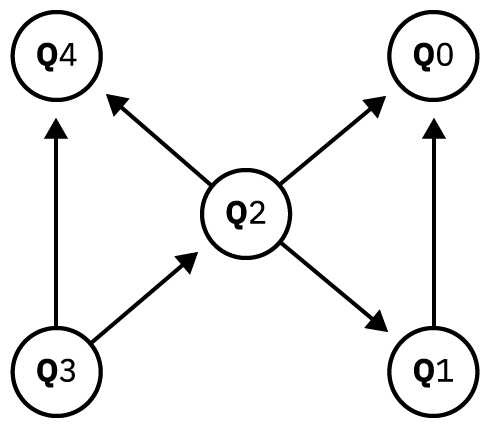}}\\
\subfigure%[$ \ $ibmqx5 ]%`Albatross']
{\includegraphics[width=0.75\linewidth]{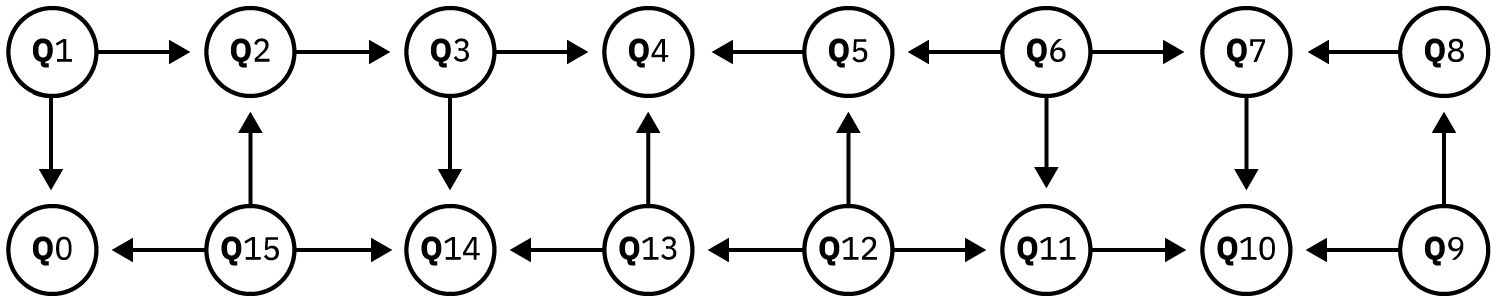}}
\caption{IBM quantum chips used in this work. Up figure corresponds to ibmqx4 and down figure to ibmqx5. Arrows indicate the directionality of CNOT gates (control {\large \ding{221}} target).
\vspace{-0.55cm}}
\label{Fig:IBM_backends}
\end{figure}

Since 2016, IBM is providing universal quantum computer prototypes based on superconducting transmon qubits which are accessible on the cloud, both interactively in their webpage (\textit{Quantum Composer})\cite{IBM} or using a software development kit (\textit{QISKit}). Currently, three quantum chips are available for the general public: two of 5 qubits, ibmqx2 and ibmqx4, and one of 16 qubits, ibmqx5 \cite{ibmbackends}. When we performed the experiments ibmqx2 was offline, so we have only used ibmqx4 and ibmqx5.

All backends work with an universal gate set composed by one-qubit rotational and phase gates
\begin{eqnarray}
U_{1}(\lambda)=\left(\begin{array}{cc}1 & 0\\0 & e^{i\lambda}\end{array}\right), \ 
U_{2}(\lambda,\phi)=\left(\begin{array}{cc} \frac{1}{\sqrt{2}} & -\frac{e^{i\lambda}}{\sqrt{2}}\\
\frac{e^{i\phi}}{\sqrt{2}} & \frac{e^{i(\lambda+\phi)}}{\sqrt{2}}\end{array}\right),\nonumber\\
U_{3}(\theta,\lambda,\phi)=\left(\begin{array}{cc} \cos(\theta/2) & -e^{i\lambda}\sin(\theta/2) \\
e^{i\phi}\sin(\theta/2)& e^{i(\lambda+\phi)}\cos(\theta/2) \end{array}\right),\nonumber\\
\label{eq:IBM}
\end{eqnarray}
and a two-qubit gate, the controlled-X or CNOT gate. 

The differences between the devices, apart from the number of qubits, come from the qubits connectivity or topology and the role that each qubit plays when applied a CNOT gate (control or target). Figure \ref{Fig:IBM_backends} shows the connectivity of the used devices. Each qubit in the 5-qubit device is connected with other two except the central one which is connected with the other four. Qubits in the 16-qubit device are connected with three neighbors in a ladder-type geometry. Both, the one-directionality of CNOT gate and the qubits connectivity, are crucial for the quantum circuit implementation. If the circuit demands an interaction between qubits that are not physically connected, we should implement SWAP gates which will increase our circuit depth and the probability of errors in our final result. Moreover, each time we need to implement a CNOT gate using as a control qubit a physical qubit which is actually a target, we have to invert the CNOT direction using Hadamard gates which, again, will increase the circuit depth and the error probability.

For our propose, ibmqx5 is the best choice for the implementation of the $n=4$ circuit. We can use any of the squares and identify upper qubits as 0 and 2 and lower qubits as 1 and 3: according to the circuit of Figure \ref{Fig:circuit},
%, qubit 0 interact with qubit 2 and 1 and qubit 3 interact with qubit 2 and 1, so 
we will not need to use any fSWAP gates. We should only take into account which qubits are control or target to try to reduce the times that we have to invert the CNOT direction.

\subsection{Rigetti Computing: Forest}

\begin{figure}[!t]
\centering
\includegraphics[width=0.85\linewidth]{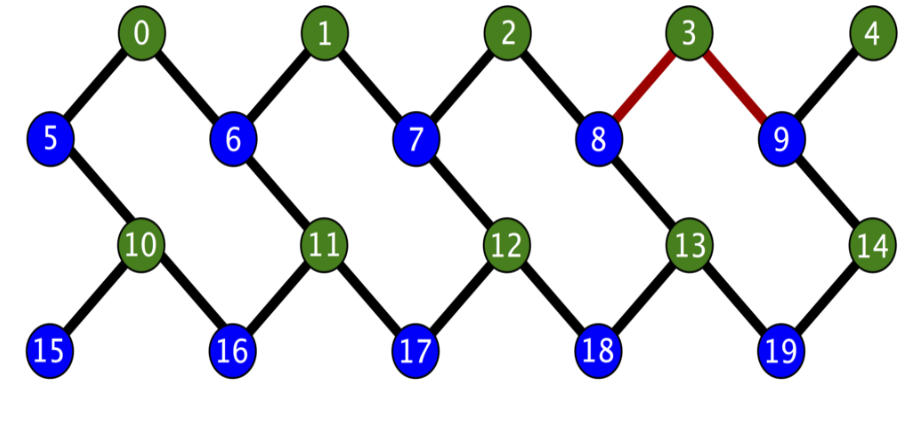}
\caption{Rigetti's 19-qubit processor `Acorn'. Lines indicate the two-qubit connection ruled by a controlled-Z gate. For technical reasons, qubit 3 is offline.\vspace*{0.2cm}}
\label{Fig:Acorn}
\end{figure}

At the end of 2017, Rigetti Computing launched a 19-qubit processor, `Acorn', that can be used in the cloud through a development environment called \emph{Forest} \cite{Rigetti}. It includes a python toolkit, \emph{pyQuil}, that allows the users to program, simulate and run quantum algorithms in a similar way as IBM's \emph{QISKit}. The chip is made of of 20 superconducting transmon qubits but for some technical reasons, qubit 3 is off-line and cannot interact with its neighbors, so it is treated as a 19-qubit device.% \footnote{See the final remarks note.}.

Currently, Rigetti's gate set is formed by three one-qubit rotational gates
\begin{equation}
\mathrm{R_{X}(\theta)}=e^{i\frac{\theta}{2}\sigma_{x}}, \
 \mathrm{R_{Y}(\theta)}=e^{-i\frac{\theta}{2}\sigma_{y}}, \ 
 \mathrm{R_{Z}(\theta)}=e^{i\frac{\theta}{2}\sigma_{z}}, 
\end{equation}
and a two-qubit gate, controlled-Z. This two-qubit gate has the advantage of bi-directionality as the result is the same independently of which is the control qubit. For that reason, the connectivity of the device shown in Figure \ref{Fig:Acorn} does not specify the direction of the two-qubit gate. 

The qubit topology is very different from IBM's devices: some qubits are connected with three neighbors and others with two in a zigzag-type geometry. 
Then, we can not do without the fSWAP gates, which means that the circuit depth will be greater than the ibmqx5's. On the other hand, it will be comparable with the ibmqx4, which also needs from these gates.

\section{Results and discussion}\label{sec:results}

\begin{figure}[t!]
\centering
\includegraphics[width=\columnwidth]{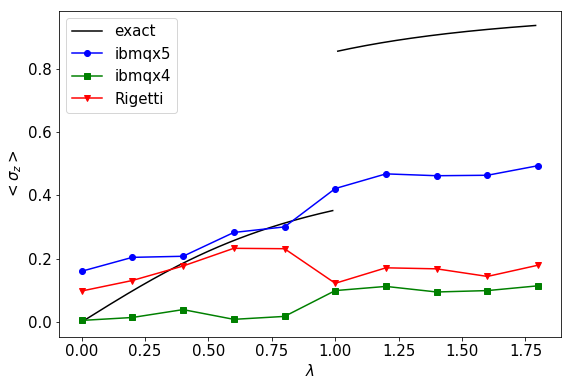}
\caption{Expected value of $\langle\sigma_z\rangle$ of the ground state of a $n=4$ Ising spin chain as a function of transverse field strength $\lambda$. Solid line represents the exact result in comparison with the experimental simulations represented by scatter points. The best simulation comes from ibmqx5 device, which is an expected result since the number of gates used is lesser than with the other devices because of qubits connectivity.
% Even though, the magnetization for $\lambda>1$ is more difficult to achieve for these devices due to any error on qubits induce a decrease in magnetization.
}
\label{Fig:mag}
\end{figure}

Figure \ref{Fig:mag} shows the results of the exact simulation of ground state transverse magnetization for the three devices. All points contain a statistical error of $1/\sqrt{N}$ with $N=1024$ which comes from the average over all runs to compute the expected value. The other error sources are discussed qualitatively in the following paragraphs.

The best performance come from the ibmqx5 device. This is an expected result as we do not need from fSWAP gates because the qubits connectivity. On the other hand, Rigetti's device performs better than the ibmqx4, even though the number of gates is very similar.

The simulation approaches better to the prediction for low $\lambda$. The explanation could come from how affect the experimental error sources to the magnetization. Assuming that two-qubit gates implementation take several hundreds of ns and single qubit gates around one hundred of ns, errors coming from decoherence are expected to be low, as these times are around 50 $\mu$s. On the other hand, errors coming from the gate implementation are cumulative and probably the most important error source. It is not negligible neither errors coming from qubits readout, which can induce a bit flip.

The analysis of the results become more clear if we look at the exact ground state wave function:
{\medmuskip=1mu
\thinmuskip=1mu
\thickmuskip=1mu
\begin{eqnarray}
|gs\rangle=\frac{1}{\mathcal{N}}
\left\{\begin{array}{l}
\alpha\left(|0001\rangle-|0010\rangle+|0100\rangle-|1000\rangle\right)\\
+|0111\rangle-|1011\rangle+|1101\rangle-|1110\rangle  \\
 \hspace{4.3cm} \mathrm{for} \ \lambda<1,\\
\alpha\left(|0011\rangle-|0110\rangle+|1001\rangle+|1100\rangle\right)\\
\hspace{2.2cm} +2|1111>  \quad \mathrm{for} \ \lambda>1, 
\end{array}\right. \nonumber\\
\label{eq:gs}
\end{eqnarray}}
where $\alpha=\lambda-\sqrt{1+\lambda^2}$ and $\mathcal{N}=2\sqrt{2}\sqrt{1+\lambda\alpha}$. As $\lambda$ increases, the amplitude for the states proportional to $\alpha$ goes to zero. That means that any error occurring for $\lambda>1$ is dramatic as it will affect the state with higher probability amplitude, the $|1111\rangle$. Then, any error in that regime will inevitably cause a decrease in magnetization. On the other hand, errors in some states for $\lambda<1$ can be compensated in average for the other elements with the same probability amplitude.

\begin{figure}[t!]
\centering
\includegraphics[width=\columnwidth]{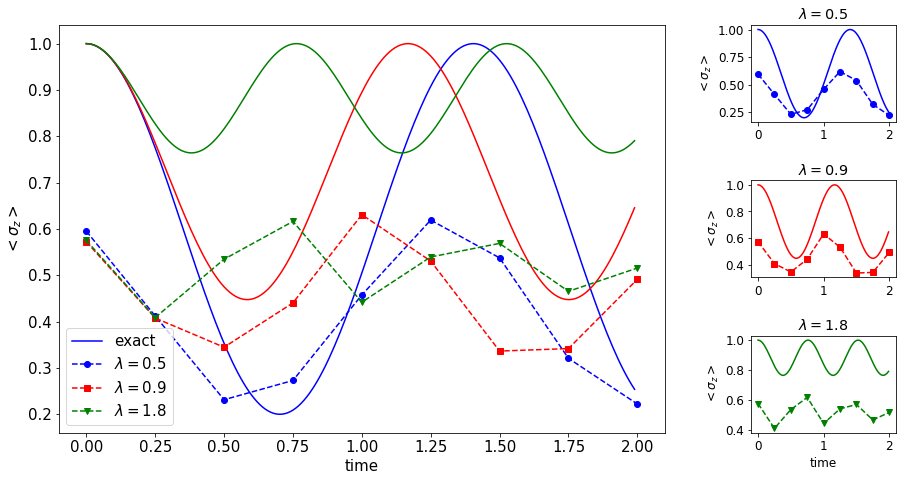}
\caption{Time evolution simulation of transverse magnetization, $\langle\sigma_z\rangle$, for the state $|\uparrow\uparrow\uparrow\uparrow\rangle$ of a $n=4$ Ising spin chain. Left plot compares the exact result with the experimental run in the ibmqx5 chip for different values of $\lambda$. Right plots detailed the results for each $\lambda$ to compare them with the theoretical values. Although the magnetization is lesser than expected, the oscillations follow the same theoretical pattern.}
%\vspace*{-0.6cm}
\label{Fig:mag_time}
\end{figure}

Similar results are obtained for the time evolution simulation. Figure \ref{Fig:mag_time} shows the results for the simulation of the $|\uparrow\uparrow\uparrow\uparrow\rangle$ state transverse magnetization as it was explained in Section \ref{sec:time}. Since for the preparation of the initial state it is necessary to implement more gates, we only show the results for the ibmqx5 device, which is the one that can afford this extra circuit depth.

As expected from the previous result, points that represent higher magnetization have more error respect the theoretical values. However, it is remarkable that the relations among the different points for different values of transverse magnetic field are proportionally correct. The oscillations take place for lower values of $\langle\sigma_{z}\rangle$, have lower amplitudes and are a little bit shifted to the left: even though, they cross each other at the corresponding points and increase and decrease proportionally to the exact result. That is a clear indicator that the error sources in the quantum device are systematic, as the result does not depend on the state preparation.

Notice that in this work we have computed the transverse magnetization instead of the staggered magnetization, i.e. $M_{x}=\sum_{i}(-1)^{i}\sigma^{x}_{i}$, which is the order parameter for the antiferromagnetic Ising model. For our purpose, it is more natural to compute $\langle\sigma_{z}\rangle$ since the states obtained with these quantum devices are expressed in the $\sigma_z$ basis. However, it will be straightforward to compute $M_{x}$ as the only change needed appears in the classical post-processing part.

\section{Conclusions}\label{sec:conclusion}
%\vspace*{-0.25cm}
In this work we have implemented the exact simulation of a one-dimensional Ising spin chain with transverse field in some quantum computers. To do so, we programmed the algorithm proposed in Ref.\cite{Latorre} in IBM and Rigetti quantum chips. We have simulated the expected value of ground state transverse magnetization as well as the time evolution of the state of all spins aligned. We have also provided two methods to compute thermal evolution of some operator using the same circuit: exact simulation or sampling.

The circuit presented allows to compute all eigenstates of the Ising Hamiltonian by just initializing the qubits in one of the states of the computational basis. It is then a implementation of a Slater determinant with a quantum computer. Since the one-dimensional Ising model is an exactly solvable model, which means that we can compute analytically all the states and energies for any number of spins, and the circuit is efficient, the number of gates scales as $n^2$ and the circuit depth as $n\log n$, it can represent a method to test quantum computing devices of any size. As has been shown, it is also a hard test since this model is strongly correlated and both the simulation of the phase transition surrounding and time evolution require a high qubits control.

The best performance has been obtained with the ibmqx5 chip, although the error respect to the theoretical prediction is large in the paramagnetic phase of the model. A possible reason why this chip shows better results than the others comes from the number of gates used in the quantum circuit, as the qubits connectivity in that device allows us to save all the fSWAP gates. On the other hand, Rigetti's chip performs better than the ibmqx4 chip, even though both implemented circuits have the same gate depth.

The paramagnetic phase is difficult to simulate due to the fact that any error that can induce a qubit bit flip will produce a decrease in magnetization, as can be traced out from the ground state wave function of Eq.\eqref{eq:gs}. However, and taking into account this fact, the time evolution simulation is reasonably good since the expected oscillations for different transverse magnetic field strengths are shifted to the left and have lower amplitude and magnetization, but are also proportional each other as are the theoretical values.

As a final remark, this circuit is also interesting from a point of view of condensed matter physics as specific methods to simulate exactly time and thermal evolution are provided. This can open the possibility of simulate other interesting models: integrable, like Kitaev Honeycomb model \cite{Orus}, or with an ansatz, like Heisenberg model.

\section*{Notes}

The program used for this work for the IBM quantum devices was awarded with the IBM ``Teach Me QISKit'' award \cite{award}.

Recently, Rigetti computing changed the quantum device to one of 8 qubits. The results shown in this work correspond to the previous device of 19 qubits.

\section*{Acknowledgements}
We acknowledge use of the IBM Q experience for this work. The views expressed are those of the authors and do not reflect the official policy or position of IBM or the IBM Q experience team. We also acknowledge use of the Rigetti computing device as well as the help and availability of Rigetti staff. This work has been possible with the support of FIS2015-69167-C2-2-P and FIS2017-89860-P (MINECO/AEI/FEDER,UE) grants. Finally, we would like to thank the discussions with Quantic group members, in particular with Jos\'e Ignacio Latorre.

%\appendix

\section*{Appendix: Gate decomposition} \label{app:gates}

%In this section we illustrate how to construct the gates needed for the implementation of $U_{dis}$ circuit.

\subsection{Fermionic-SWAP}
\begin{figure}[h!]
\[
\Qcircuit @C=1em @R=1em @!R {
& \qswap & \ctrl{1} & \qw & \push{\rule{0em}{0em} \rule{0em}{0em}} & & \ctrl{1} &  \targ  & \ctrl{1} & \qw & \ctrl{1} & \qw & \qw \\
 & \qswap \qwx & \gate{Z} & \qw & \push{\rule{0em}{0em}\equiv\rule{0em}{0em}}&  & \targ  & \ctrl{-1} & \targ & \gate{H} & \targ & \gate{H}& \qw
}
\]
\caption{Fermionic SWAP gate. 
%It consist in a SWAP gate an a controlled-Z gate (CZ): SWAP gates can be decomposed into three CNOT gates and CZ gate is equivalent to a Hadamard-CNOT-Hadamard gates as shown in figure.
}
\label{Fig:fSWAP}
\end{figure}
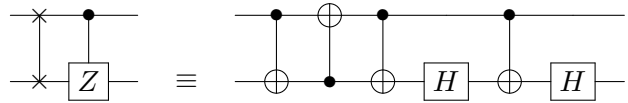

The Jordan-Wigner transformation do not need from any quantum gate, but as it transforms the spin operators $\sigma$ into fermionic modes $c$, any swap between qubits should obey the fermionic anticommutation relations, i.e. the exchange between two occupied modes carries a minus sign. This is represented with the use of fermionic SWAP gate (fSWAP), decomposed in basic gates in Figure \ref{Fig:fSWAP}.

From the point of view of IBM's implementation, at least one CNOT should be inverted to fit the circuit to the qubits connectivity; this can be easily done using the identity $(H_{1}\otimes H_{2})\overrightarrow{\mathrm{CNOT}}(H_{1}\otimes H_{2})=\overleftarrow{\mathrm{CNOT}}$. In addition, controlled-Z gate could be implemented using two Hadamard gates and a CNOT, as it is also shown in Figure \ref{Fig:fSWAP}.

From Rigetti's implementation point of view, controlled-Z gates are part of their basic gate set and, although CNOT gate and H are not, they are included in $pyquil$ language, so the quantum programmer should not care about decompose them in terms of the other gates (except to keep in mind the circuit depth will increase with the use of non-basic gates).

\subsection{Fourier transform}

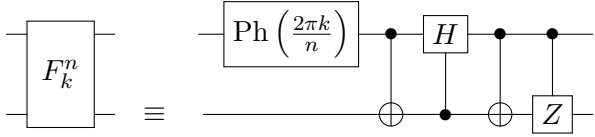
\begin{figure}[h!]
\[
\Qcircuit @C=0.7em @R=1em %@!R 
{
&\multigate{1}{F_{k}^{n}}& \qw &\push{\rule{0em}{0em} \quad\quad \rule{0em}{0em}}& \gate{\mathrm{Ph}\left(\frac{2\pi k}{n}\right)} & \ctrl{1} & \gate{H} & \ctrl{1} & \ctrl{1} & \qw \\
&\ghost{F_{k}^{n}}& \qw& \push{\rule{0em}{0em}\equiv\quad\rule{0em}{0em}} & \qw &\targ & \ctrl{-1} & \targ & \gate{Z} & \qw 
}
\]
\caption{Decomposition of the building block of Fourier transform gate. The controlled-Hadamard gate is shown in Figure \ref{Fig:CH}. 
%Phase gate $\mathrm{Ph}$ is included in Rigetti's $quil$ language and it is equivalent to $U_{1}$ gate in the IBM's devices.
}
\label{Fig:Fk}
\end{figure}

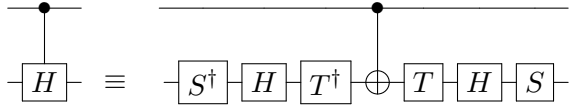
\begin{figure}[h!]
\[
\Qcircuit @C=0.5em @R=1.7em %@!C 
{
& &\ctrl{1} & \qw &\push{\rule{0em}{0em} \quad\quad \rule{0em}{0em}}& \qw & \qw & \qw & \ctrl{1} & \qw & \qw & \qw & \qw \\
& & \gate{H} & \qw & \push{\rule{0em}{0em}\equiv\quad\rule{0em}{0em}}&\gate{S^{\dagger}} & \gate{H} & \gate{T^{\dagger}} & \targ & \gate{T} & \gate{H} & \gate{S} & \qw
}
\]
\caption{Controlled-Hadamard gate. 
%Gates $S$ and $T$ are included in IBM's basic gate set and correspond to $\mathrm{Ph}(\pi/2)$ and $\mathrm{Ph}(\pi/4)$ respectively.
}
\label{Fig:CH}
\end{figure}

The building blocks of Fourier transform gate are represented by the quantum gate of Eq.\eqref{F_gate}, which decomposition is shown in Figure \ref{Fig:Fk}. Its implementation requires the controlled-Hadamard gate, decomposed in Figure \ref{Fig:CH}. Phase gate $\mathrm{Ph}$ is included in Rigetti's $pyquil$ language and it is equivalent to IBM's $U_1$ gate (see Eq. \eqref{eq:IBM}). In particular, $\mathrm{Ph}(\pi/2)$ and $\mathrm{Ph}(\pi/4)$ correspond to $S$ and $T$ gates respectively.\\

\vspace*{-0.6cm}

\newpage

\subsection{Bogoliubov transformation}

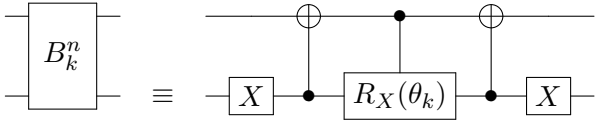
\begin{figure}[h!]
\[
\Qcircuit @C=0.8em @R=1.5em %@!R 
{
&\multigate{1}{B_{k}^{n}} & \qw & \push{\rule{0em}{0em}\quad \ \rule{0em}{0em}}& & \qw & \targ & \ctrl{1} & \targ & \qw & \qw \\
&\ghost{B_{k}^{n}} & \qw & \push{\rule{0em}{0em}\equiv\rule{0em}{0em}} & & \gate{X} &  \ctrl{-1} & \gate{R_{X}(\theta_{k})} & \ctrl{-1} & \gate{X} & \qw
}
\]
\caption{Bogoliubov gate decomposition. Controlled-$RX$ gate needed is shown in Figure \ref{Fig:RX}.}
\label{Fig:Bk}
\end{figure}

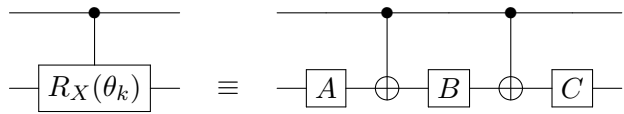
\begin{figure}[H]
\[
\Qcircuit @C=1em @R=1em @!R {
&\ctrl{1} & \qw & \push{\rule{0em}{0em} \rule{0em}{0em}} & & \qw & \ctrl{1} & \qw & \ctrl{1} & \qw & \qw \\
&\gate{R_{X}(\theta_{k})} & \qw & \push{\rule{0em}{0em}\equiv\rule{0em}{0em}}&  & \gate{A} & \targ & \gate{B} & \targ & \gate{C} & \qw
}
\]
\caption{Controlled-$RX$ gate decomposition in terms of the rotational gates $A=R_{Z}\left(\frac{\pi}{2}\right)R_{Y}\left(\frac{\theta}{2}\right)$, $B=R_{Y}\left(-\frac{\theta}{2}\right)$ and $C=R_{Z}\left(-\frac{\pi}{2}\right)$.}
\label{Fig:RX}
\end{figure}

Bogoliubov transformation is implemented using $B_{k}^{n}$ gates written in Eq. \eqref{B_gate}. The explicit decomposition is shown in Figure \ref{Fig:Bk}, where the controlled-$RX$ gate (shown in Figure \ref{Fig:RX} has been decomposed using the methods of Ref.\cite{Barenco}). Rotational gates are part of the basic Rigetti gate set and are equivalent to IBM's gates $R_{X}\equiv U_{3}(\phi=0,\lambda=\pi)$, $R_{Y}\equiv U_{3}(\phi=\lambda=0)$ and $R_{Z}\equiv U_{1}$.

\vspace*{-1cm}
\vspace*{-0.5cm}
\subsection{Initial state preparation}

The ground state of the $n=4$ Ising model in the diagonal basis is $|0000\rangle$ for $\lambda>1$ and $|0001\rangle$ for $\lambda<1$. Qubits are always initialized in the $|0\rangle$ state both in IBM and Rigetti devices. Then, to compute the ground state, we only need to perform a bit-flip gate (Pauli-X or $X$ gate), on fourth qubit to initialize the circuit for $\lambda<1$.

The example given for time evolution simulation requires from the preparation as the initial state the one shown in Eq.\eqref{eq:time_Ising}. This can be done by applying a $R_{Y}(\phi)$ gate on the first qubit to introduce the $\phi$ angle, followed by a phase gate to introduce the evolution phase $e^{4it\sqrt{1+\lambda^2}}$ and a CNOT gate between first and second qubits.

%\vspace*{2.5cm}

\end{document}